# Exploring More-Coherent Quantum Annealing

Sergey Novikov[1], Robert Hinkey[1], Steven Disseler[1], James I. Basham[1], Tameem Albash[2,3,4], Andrew Risinger[1], David Ferguson[1],
Daniel A. Lidar[3,4,5,6], Kenneth M. Zick[1]

[1]Northrop Grumman Corporation, Linthicum, MD 21090 USA
sergey.novikov@ngc.com
[2]Information Sciences Institute, University of Southern California, Marina del Rey, CA 90292 USA
[3]Department of Physics and Astronomy, University of Southern California, Los Angeles, California 90089 USA
[4]Center for Quantum Information Science & Technology, University of Southern California, Los Angeles, California 90089 USA
[5]Department of Electrical Engineering, University of Southern California, Los Angeles, California 90089, USA
[6]Department of Chemistry, University of Southern California, Los Angeles, California 90089, USA
{albash, lidar}@usc.edu

*Abstract*—In the quest to reboot computing, quantum annealing (QA) is an interesting candidate for a new capability. While it has not demonstrated an advantage over classical computing on a real-world application, many important regions of the QA design space have yet to be explored. In IARPA's Quantum Enhanced Optimization (QEO) program, we have opened some new lines of inquiry to get to the heart of QA, and are designing testbed superconducting circuits and conducting key experiments. In this paper, we discuss recent experimental progress related to one of the key design dimensions: qubit coherence. Using MIT Lincoln Laboratory's qubit fabrication process and extending recent progress in flux qubits, we are implementing and measuring QA-capable flux qubits. Achieving high coherence in a QA context presents significant new engineering challenges. We report on techniques and preliminary measurement results addressing two of the challenges: crosstalk calibration and qubit readout. This groundwork enables exploration of other promising features and provides a path to understanding the physics and the viability of quantum annealing as a computing resource.

*Keywords—quantum annealing, superconducting, qubit, coherence, quantum computing*

## I. Introduction

In the quest to reboot computing, one interesting candidate for a new capability is quantum annealing (QA). Its origins trace back to the 1980s and 1990s, first as a quantum-inspired classical optimization method and eventually as a proposed type of analog quantum computing [1]. The concept of operation typically involves initializing a carefully isolated quantum system, applying analog controls such that the energy landscape comes to represent a user's classical problem of interest (the annealing step), then reading out the final state. The final state is meant to be a low-energy state corresponding to a good solution to the original problem. The intuition is that quantum mechanics provides unique phenomena that enhance the ability of a system to navigate a complex energy landscape and find those low-energy states. These solutions can prove very valuable if they are much harder (or intractable) to find using classical computing. Such a capability would have high impact in discrete combinatorial optimization or potentially in sampling applications such as those found in machine learning.

For combinatorial optimization, there are currently three heuristic quantum approaches: QA (analog), a digital variant of annealing (Trotterized) [2], and a hybrid approach called the quantum approximate optimization algorithm (QAOA) [3] that would run on a gate-model capable quantum computer. In classical computing, digital computers ultimately supplanted analog computers; thus, one could ask why one would explore QA rather than a digital-capable device. One observation is that classical analog devices proved to be very useful for an extended period, and may yet return to prominence in novel computing architectures such as neuromorphic computing [4]. Second, QA might have an advantage of being more robust to certain forms of decoherence like thermal relaxation [5], since the evolution of the system primarily follows the ground state. Moreover, because the energy landscape is changed relatively slowly, the control requirements for QA should be significantly less demanding. For instance, extensive use of microwave technology for qubit control may not be required. It might be possible in the near term to engineer QA systems with sufficiently low noise such that they outperform digital quantum devices subject to higher overhead costs.

Currently, there is no evidence that QA can show a scaling advantage (e.g. a polynomial or exponential speedup) for solving optimization problems. On the other hand, there is also no theorem that *excludes* a scaling advantage for QA. In many cases, new computing technologies achieve practical utility without theoretical proof of a scaling advantage (deep learning is one such example). Additional impetus to investigate QA as a computational resource comes from the fact that many of the promising regions of the design space have not yet been explored. One particularly important dimension in this space is coherence. The coherence of a quantum system is a time scale for which we can expect the system to evolve in a quantum-mechanical fashion. Without coherent evolution, no quantum

This material is based upon work supported by the Intelligence Advanced Research Projects Activity (IARPA) through the Army Research Office (ARO) under Contract No. W911NF-17-C-0050. Any opinions, findings and conclusions or recommendations expressed in this material are those of the author(s) and do not necessarily reflect the views of the Intelligence Advanced Research Projects Activity (IARPA) and the Army Research Office (ARO). Approved For Public Release #18-1898; Unlimited Distribution, Dated 09/06/18.

advantage (if it exists) can be expected. Coherence is often characterized by two time scales, the $T_1$ relaxation time and the $T_2$ dephasing time of the qubit [6]. While these decoherence time scales are a good indication of how isolated the qubit is from its environment, they do not provide a complete picture of decoherence in QA; other considerations will be covered in Section II.

A natural qubit technology for implementing QA is tunable superconducting flux qubits [7][8]. The states of such a qubit are characterized by currents flowing in opposite directions in the circuit's main loop. The direction of these currents maps onto the binary spin variable used in QA. Further, the magnitude of the qubit current is tunable and can be controlled via an applied magnetic flux. During annealing operation the qubit currents are made to dramatically increase in magnitude, arriving (at the end of the anneal) in one of two persistent current states. The direction of the final persistent current depends on the effective bias due to the qubit's interaction with other qubits. At the end of the anneal, this current is measured for every qubit to obtain the bit-string corresponding to the solution. D-Wave Systems has conducted interesting experiments and done pioneering work toward QA [9][10], however the technology described to date relies on very high persistent currents [9] which generally implies very low coherence. To our knowledge, measurement data showing qubit coherence times of D-Wave processor qubits have never been published. Recently, a capacitively-shunted flux qubit (CSFQ) style was developed by MIT Lincoln Laboratory and the University of California, Berkeley that demonstrated a path to greatly increased coherence. Though not in the direct context of QA, those flux qubits achieved reproducible coherence times over an order of magnitude longer than previous state-of-the-art [11][12]. Other efforts toward more-coherent QA are underway [13].

Many interesting questions about QA remain, starting with the question of what can be achieved with more-coherent qubits. Such questions will likely need to be addressed experimentally. A concise summary of the state of affairs was recently provided by John Preskill, who said, "Since theorists have not settled whether quantum annealing is powerful, further experiments are needed" [14].

As part of IARPA's Quantum Enhanced Optimization (QEO) program, we have opened promising new lines of inquiry in an attempt to get to the heart of QA. In this paper, we discuss recent experimental progress related to qubit coherence. Such work is a necessary precursor to experimental explorations of other key features of interest. First, in Section II we discuss our recent work designing and conducting measurements of annealing-capable flux qubits, building on MIT Lincoln Laboratory's CSFQ style and qubit fabrication process [11][12]. Our targeted regime with higher coherence presents unique engineering challenges. Section III covers the challenge of crosstalk and our progress in developing and validating a technique for calibration. Section IV covers qubit readout, including initial designs and measurement results showing progress toward high-fidelity readout capability. Finally, in Section V we conclude with some thoughts about the implications of this work and the outlook for exploring quantum annealing.

## II. QUBIT COHERENCE

Loosely speaking, when we refer to the "coherence" of a system of qubits we mean how long we can expect the system's evolution to be primarily dictated by quantum mechanics (specifically, the unitary dynamics of the Schrödinger equation), before the effect of the environment becomes non-negligible. $T_1$ and $T_2$ times are measured with respect to a fixed Hamiltonian. (The Hamiltonian is the operator that generates the time evolution of the system and in most cases corresponds to the operator associated with the total energy of the system.) However, in QA the Hamiltonian is constantly changing during the annealing process, so $T_1$ and $T_2$ times give only partial information about coherence and decoherence. It is crucial to unambiguously determine the basis in which the decoherence is occurring. At any point during the anneal, the eigenvalues and eigenstates of the instantaneous Hamiltonian define the instantaneous spectrum. Dephasing in the instantaneous energy eigenbasis (corresponding to the loss of the relative phase between energy eigenstates in the quantum state) is innocuous for QA but dephasing in the computational basis is crippling [5]. In the most favorable decoherence model (in which the system is weakly coupled to a Markovian environment), the dephasing occurs in the instantaneous energy eigenbasis and the only relevant decoherence is relaxation to the thermal state. Because the thermal state can have population not only in the ground state but in excited states as well, thermal relaxation typically depopulates the ground state. Thermal effects become especially problematic when 1) the energy gap between the ground state (which tracks the desired solution during the computation) and the excited states becomes of the same order as the temperature of the environment, and 2) the annealing time is long enough compared to $T_1$ to make the probability of an excitation significant.

One of the key requirements in QEO is the development of QA-capable qubits that are sufficiently robust against the relevant decoherence mechanisms, thereby pushing deep into the coherent regime. A key driver of decoherence in superconducting flux qubits is the amount of persistent current, which is defined as the amount of current flowing in the main ("Z") loop of the qubit circuit. It is chiefly determined by circuit parameters such as the qubit's Josephson junction critical currents. Currently available commercial annealers developed by D-Wave Systems utilize niobium-based flux qubits with large persistent currents ($I_p \sim 3~\mu A$) [9] that aid in coupling qubits strongly together but lead to short coherence times.

The increased coherence of MIT Lincoln Laboratory's recent CSFQs was due to several factors. The foremost is the reduced amount of persistent current ($I_p$) in the qubit loops. $I_p$ determines a qubit's sensitivity to flux noise, with coherence times scaling approximately as $T_1 \sim \frac{1}{(I_p)^2}$ and $T_2 \sim \frac{1}{I_p}$. The amount of persistent current is perhaps the most important design decision for creating highly-coherent flux qubits. CSFQ experiments achieved persistent currents much lower than before, with $I_p \sim 50\ nA$. The second important factor is the increased capacitance that shunts the qubit's Josephson junctions. This feature reduces the sensitivity to charge noise and improves device reproducibility. Thirdly, fabrication of the qubit shunt capacitors using high-quality aluminum on a low-loss substrate reduces the number of material defects that could degrade coherence. The last factor is the use of shunt capacitors that are physically large, a technique recently pioneered in another qubit technology [15]. This design choice reduces the amount of a qubit's electric field energy stored in lossy interfaces, thereby further increasing the qubit's $T_1$ energy relaxation time.

In order to investigate the role of coherence in QA, QEO is conducting annealing experiments that leverage MIT Lincoln Laboratory's qubit fabrication process. In this hybrid process, qubit and coupler loops are patterned with aluminum using a shadow-evaporated technique, while capacitive shunts, control circuitry, and readout circuitry are patterned with aluminum using a different technique (molecular beam epitaxy). While this fabrication process maximizes coherence, it also results in increased unwanted crosstalk between control lines on the chip. This is because all control lines exist in the same plane, and in contrast to a multi-layer fabrication process, there are no ground/sky planes or via walls to contain the electromagnetic fields. We have implemented initial solutions to calibrate the crosstalk, described in detail in the following section. As our systems grow in size, careful design of both the chip and package will be required to minimize the crosstalk, as well as to constrain its locality.

The annealing process consists of dynamically changing control bias signals to evolve the energy landscape into one representing the problem of interest. After the annealing is complete, the output of the computation is extracted by reading out the state of each qubit. The strength of the signal available for readout is proportional to the qubit persistent current. As mentioned above, the high-coherence CSFQ parameter regime necessitates a small $I_p$, which presents a challenge for high-fidelity readout. In Section IV, we show how we have overcome this challenge, and present a path for further performance improvements.

Building upon the previous CSFQ work, we have designed and measured CSFQs that are now capable of being annealed (see Fig. 1a for the schematic of the qubit) and coupled to multiple other devices in a QA context. These qubits have an inner loop called an X loop (corresponding to the X axis of the Bloch sphere). The presence of this loop makes the qubit annealing-compatible: the magnetic flux $\Phi_x$ through the X loop can be controlled to enact the annealing protocol. There is also an outer loop called the Z loop; the magnetic flux $\Phi_z$ through this loop is used to encode a variable in the problem of interest. The measured qubit was designed to have Josephson junction critical currents of 90 nA (X loop) and 186 nA (Z loop), and a shunt capacitance of 45 fF. We have conducted preliminary experiments to measure the coherence of these tunable qubits. Characterization measurements of a prototype qubit are shown in Fig. 1. The energy relaxation time $T_1$ was acquired using the standard method of measurement of the excited state population as a function of time following a single $\pi$ pulse to invert the population from ground to excited state. As shown in Fig. 1b, we measured $T_1 = 3.5\ \mu s$. This is an encouraging result for annealing-compatible flux qubits, and compares favorably to recent results from other groups [13]. We were also able to obtain preliminary dephasing time measurements, another early sign of promise for annealing-compatible flux qubits. The dephasing time $T_2^*$ was acquired by Ramsey free-induction decay where the driving pulse was detuned by several MHz from the qubit transition. Fig. 1c shows the decay of Ramsey oscillations of the qubit, from which we extract $T_2^* = 130\ ns$. Ramsey decay is a result of low-frequency noise, and $T_2^*$ provides a lower bound for the qubit's $T_2$ dephasing time ($T_2 > T_2^*$). We note that during these measurements, the qubit was biased at a typical operating point ($\Phi_x = 670\ m\Phi_0$), and therefore our measurements are indicative of the realistic rather than best-case performance. Over the course of the QEO program we will continue to optimize the noise environment and the noise sensitivity of the qubits themselves, and improve the electromagnetic environment engineering to minimize direct qubit coupling to spurious device package modes.

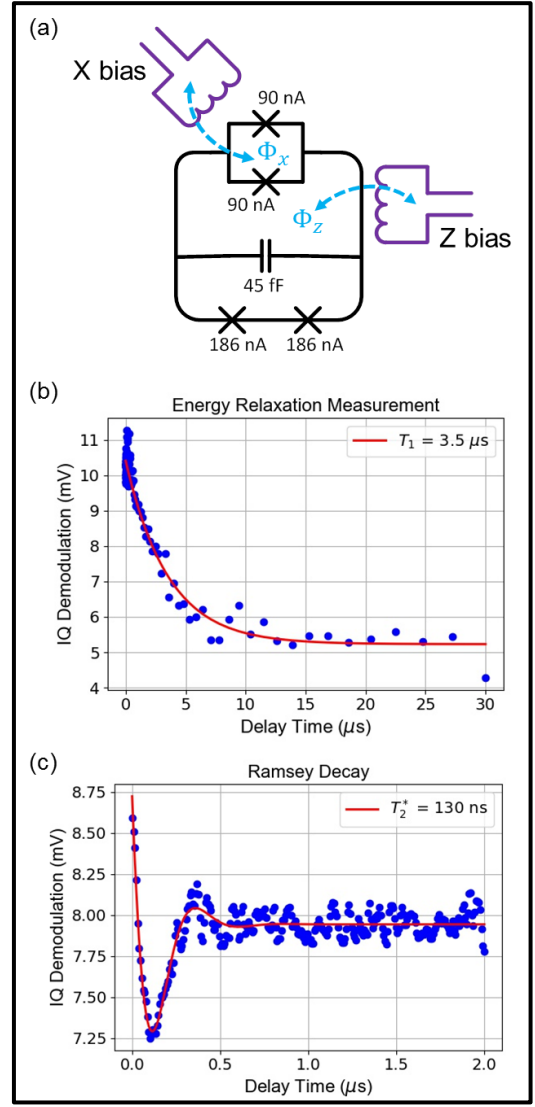

Fig. 1. (a) Schematic of annealing-compatible C-shunt flux qubit. Control lines are shown in purple, with the corresponding flux to the qubit's X and Z loops in blue. Measurement data (blue dots) and theory fits (red lines) showing (b) energy relaxation and (c) dephasing of the device. For the measurements reported here, the X loop flux bias was set to $\Phi_x = 670\ m\Phi_0$, where $\Phi_0$ is the magnetic flux quantum, and the Z loop flux was biased near the flux-insensitive point of zero flux, resulting in the qubit frequency $f_{01} = 4.2$ GHz.

The presented $T_1$ and $T_2^*$ are nevertheless sufficient to start exploring the role of coherence in QA. In this regard, we are pursuing an approach of "stress" testing our qubit systems by deliberately searching for annealing problem instances with as-small-as-possible minimum energy gaps. We believe these "small-gap" problems can provide valuable insight into the role of coherence even at small system sizes. The first such system we are investigating is a 3-qubit triangle (a $K_3$ graph with pair-wise connected qubits). As we increase the number of qubits in our devices, other problems with even smaller gaps will become accessible, providing additional insight into the role of coherence, and its interplay with annealing speed, problem hardness, and device temperature.

## III. Crosstalk Calibration

Accurately encoding a classical problem of interest on a quantum annealer requires precise control of the circuit elements. In QEO devices, bias control lines are used to control the applied flux in each of the X and Z loops of qubits, couplers, and readout resonators. Ideally, this provides full and independent control over each of these parameters. However, in reality there is significant crosstalk between various control lines, which leads to unintended interference of control signals [16]. This "geometric" crosstalk is the result of unscreened magnetic fields generated by control currents coupling via mutual inductance into unwanted parts of the circuit. The QEO fabrication process for die containing qubits provides a single metal layer, barring us from using standard multi-layer crosstalk shielding techniques such as ground/sky planes and via walls. The QEO fabrication roadmap includes 3D integration of die on multiple tiers; this will allow us to implement appropriate practices to reduce crosstalk, but some crosstalk will inevitably remain, necessitating correction methods.

For each qubit or coupler we typically have at least two control lines – one to provide flux $\Phi_x$ to the X loop, and one to provide flux $\Phi_z$ to the Z loop, shown in Fig. 2a. Due to physical proximity, the crosstalk between two control lines of a qubit (or a coupler) is likely to be the dominant type. Crosstalk between adjacent qubits and couplers is smaller but non-negligible. One can define a control matrix $\vec{\vec{M}}$ which relates a vector of the control currents $\vec{I}$ to the resultant vector of magnetic fluxes $\vec{\Phi}$ biasing various device loops: $\vec{\Phi} = \vec{\vec{M}}\vec{I}$. In this form, the designed control signals are the on-diagonal elements of $\vec{\vec{M}}$, while off-diagonal elements represent crosstalk. To compensate for the crosstalk, one must find a transformation matrix $\vec{\vec{T}}$ that diagonalizes $\vec{\vec{M}}$. Then, $\vec{\vec{M}}' = \vec{\vec{T}}^{-1}\vec{\vec{M}}\vec{\vec{T}}$ where $\vec{\vec{M}}'$ is diagonal. This yields a corrected flux control of $\vec{\Phi'} = \vec{\vec{M}}'\vec{\vec{T}}^{-1}\vec{I}$. Typically, $\vec{\vec{M}}$ has a block-diagonal form because the crosstalk is only significant when bias control lines are in close proximity to each other, with each block of $\vec{\vec{M}}$ describing crosstalk in the few adjacent control lines. This limits the dimensionality of the crosstalk matrix even as the number of qubits is scaled up. Given this, we are optimistic that crosstalk calibration complexity will not increase significantly as larger systems are built.

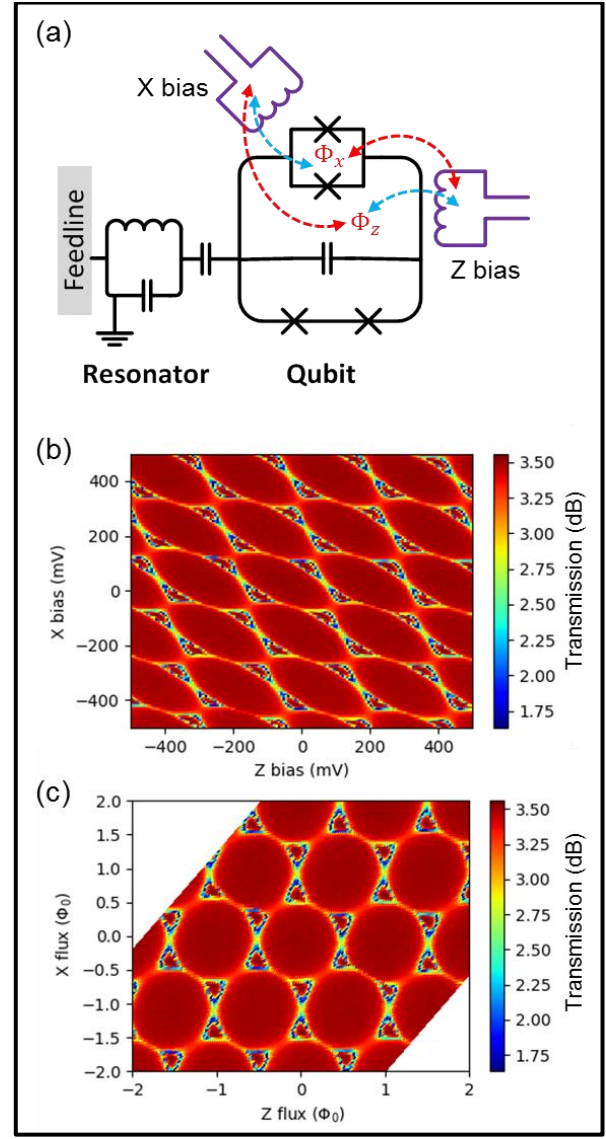

Fig. 2. (a) Device schematic showing the qubit, its readout resonator, and bias control lines (purple). The control lines apply both desired (blue arrows) and undesired (red arrows) flux into the X and Z loops of the qubit. (b) Color plot of resonator transmission versus X and Z biases showing significant crosstalk. (c) Resonator transmission versus biases after crosstalk correction has been applied to the qubit controls.

We have developed and implemented a crosstalk calibration technique, which we demonstrate here by applying it to a single-qubit circuit. First, we step the nominal qubit X and Z flux biases while monitoring transmission at a single frequency through a readout resonator attached to the qubit. This process generates a two-dimensional figure, shown in Fig.

2b. Note the significant crosstalk: the system is periodic in both dimensions but the periodicity axes are not aligned with the X and Z control flux axes. To generate the correction transformation matrix, we manually identify the center coordinates of the "hexagons" in the uncorrected crosstalk data of Fig. 2b, and use an algorithm that calculates the diagonalizing matrix $\overleftrightarrow{T}$ from these coordinates [17]. The inverse of this affine transformation, $\overleftrightarrow{T}^{-1}$, is then applied to the qubit control instrumentation to correct for the crosstalk ($\vec{I} \to \overleftrightarrow{T}^{-1}\vec{I}$) as well as any flux offsets due to stray magnetic fields or trapped magnetic flux within the chip. With these corrections in place, we obtain the data shown in Fig. 2c. One can see that X and Z flux axes are now aligned with the axes of periodicity, indicating that we have the desired orthogonal control over the qubit biases.

While this scheme works for small numbers of bias lines, it does not scale favorably to the large number of qubits and couplers planned for future experiments. In fact, the measurement time scales exponentially as $O(t^N)$ for $N$ interacting bias lines. Locality of the crosstalk is the only potential saving grace, limiting $N$ to a small number. Considering that the data shown in Fig. 2b was acquired in approximately 3 hours for just two control lines, we can extrapolate that calibration in this manner for a complex device with several qubits and couplers would require thousands of hours – an intractable task. To overcome this obstacle, we have developed a method inspired by the work in the spin-qubit community [18][19] which leads to massive data acquisition speedup. We use arbitrary waveform generators to apply saw-tooth signals to the control lines of interest, at a frequency of 100 Hz to 1 kHz. The resonator is then sampled to achieve the desired resolution along each bias ramp, with the response measured using a microwave autodyne technique [20] for 1-5 µs at each sample point. The resulting signal for many ramp cycles can then be acquired, stored, and averaged on a digitizer card, meaning at least one control dimension can be acquired on the millisecond time scale. As a proof-of-principle, we used this method to re-take the data in Fig. 2b. Compared to the original 3 hours, our new method acquired similar data in 10 seconds, yielding a speedup of over 1000x.

The reliance on user input to generate a correction matrix is a challenge to scalability, due to the difficulty in visualizing and recognizing patterns in data with $N > 2$ dimensions. We are currently developing a method to reduce the number of input data points needed for crosstalk correction. We are also investigating increasingly automated calibration using computer vision techniques. Together with the fast data collection, this will enable efficient and hands-off calibration applicable to larger-scale systems.

## IV. QUBIT READOUT

As with other forms of quantum computing, one of the challenges in quantum annealing is reading out the answer after the computation is completed. At the end of an annealing cycle, each flux qubit will be in a well-defined persistent current state. Current in the qubit's Z loop will be circulating either clockwise or counterclockwise, forming an effective magnetic dipole pointing down (state $|\downarrow\rangle$) or up (state $|\uparrow\rangle$), respectively. The bit string consisting of each qubit state represents the result of the annealer's computation. To read the state of a qubit, one can use a small magnetometer located near the qubit loop to measure the sign of the magnetic field (i.e. the direction of current in the qubit loop). It is common to use a SQUID for this purpose [21][22][23], as it can be highly sensitive, with sensitivity that can be varied in-situ by applying flux to the SQUID loop. As described in Section II, the high coherence of CSFQs is achieved through low qubit persistent currents. Our typical annealing-capable flux qubits have $I_p \sim 100\ nA$. Typical mutual inductances $M \sim 50\ pH$ imply one has to be able to measure $\Phi = I_p \times M \sim 5 \times 10^{-18}\ Wb \sim 2\ m\Phi_0$ of magnetic flux. SQUID circuits can provide the high sensitivity required to measure such small amounts of flux.

In the annealed state, the qubit is in a strong double-well potential regime, where the inter-well barrier is high enough that the tunneling rate between the two wells is negligibly small [24]. Furthermore, the phase relationship between the computational basis states is lost. As such, the qubit state after the anneal is described by a classical quantity – the persistent current in its Z loop – and the readout is a deterministic measurement of this current. Longer readout time leads to a better signal-to-noise ratio, and a lower chance of readout error. On the other hand, fast readout is desired to minimize the overall time per annealer run.

We have developed a readout scheme that includes a resonator terminated by an rf-SQUID inductively coupled to the qubit (see Fig. 3a). Magnetic flux due to the qubit's persistent current state threads the SQUID loop; the sign of this flux affects the SQUID's effective inductance which in turn determines the resonator's frequency. At the point where the qubit persistent current changes sign (a transition from the $|\downarrow\rangle$ to the $|\uparrow\rangle$ state), there is a discontinuous jump in the resonant frequency (Fig. 3b). The difference in microwave transmission through the resonator at a fixed frequency near this jump is what determines the readout contrast. Our device displays a shift of 15 MHz (14 line widths) enabling a high-contrast measurement. To read out, one measures the transmitted power of an injected single-frequency microwave tone matching the expected resonator frequency. In the example of Fig. 3 we chose to measure the transmitted power at the lower edge of the transition, $f = 6.003$ GHz. When the qubit is in the $|\downarrow\rangle$ state, the resonator frequency matches the frequency of the probe microwave tone, resulting in low transmission due to the resonator absorbing the probe tone. Conversely, if the transmission is high, the resonant frequency must be shifted, implying the qubit is in the $|\uparrow\rangle$ state.

Readout is performed as a single-shot measurement: for each experimental trial, the qubit is annealed then read out once. An ensemble of such measurements is made, and the data binned in a histogram, typically resulting in a bimodal distribution (Fig. 3c). The two Gaussians in the histogram correspond to the $|\downarrow\rangle$ and $|\uparrow\rangle$ states of the qubit. Large separation between the two Gaussians is therefore necessary to discriminate between states. A threshold value for the readout voltage (transmission value) is set to define the $|\downarrow\rangle$ state (below threshold) and the $|\uparrow\rangle$ state (above threshold). We built a proof of concept circuit that has a high-Q resonator ($Q \approx 10{,}000$) and used a long readout time (10 μs) to demonstrate histograms with centers separated by 11 standard deviations, and a measured readout fidelity of >99.99% (Fig. 3c). Readout fidelity in the QA context is the measure of how successfully one can read out a state that was prepared via annealing (e.g. the probability of successfully reading out $|\uparrow\rangle$ after preparing $|\uparrow\rangle$).

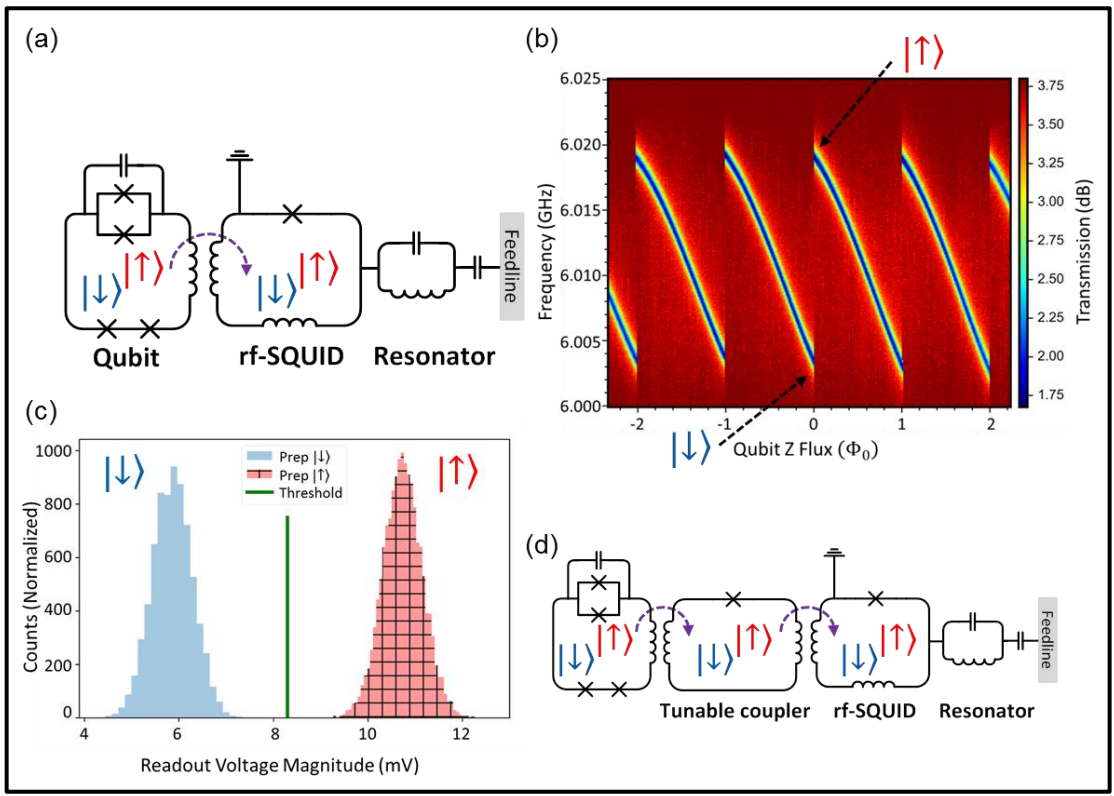

Fig. 3. (a) Schematic of the measured device demonstrating high-fidelity readout. (b) Resonator transmission data versus probe frequency and qubit Z bias. The qubit is prepared in either the $|\downarrow\rangle$ state for biases just less than $n\Phi_0$ or the $|\uparrow\rangle$ state for biases just larger than $n\Phi_0$. The discrete jumps in the resonant frequency (low transmission) occur when the qubit state switches. The jumps are 14 resonator linewidths in magnitude, enabling high-contrast readout. (c) Single-shot measurement of resonator transmission at $f = 6.003$ GHz for the qubit prepared in the $|\uparrow\rangle$ or the $|\downarrow\rangle$ state, showing $11\sigma$ of separation or fidelity of >99.99%. (d) Improved readout circuit, allowing isolation of the qubit from the lossy resonator during the anneal, and a fast readout when the computation is done.

Having proved the viability of this initial design, a potential next step would be to lower the resonator Q value to enable a fast ring-up of the resonator and shorter readout times. A low Q resonator would normally allow greater environmental noise to couple to the qubit, with a negative impact on qubit coherence. Inserting a tunable coupler between the qubit and the resonator (Fig. 3d) could allow one to isolate the qubit from the resonator during the computation (when high coherence is needed), while coupling the two together for fast readout after the anneal.

## V. Discussion and Conclusion

Just as computing is ripe for a reboot, so too is quantum annealing. One of the fundamental features to explore experimentally is high coherence. In this work, we have demonstrated qubits that are compatible with QA and at the same time capable of much higher coherence than those in current annealers. Moreover, we have identified some of the unique engineering challenges associated with this region of the design space, and have implemented and validated initial techniques for both crosstalk calibration and qubit readout. Such experimental work is a necessary precursor to more advanced experiments. Building upon this technology base, there is now a path to experiments involving other key features. One line of inquiry will explore coupling of multiple qubits together in a larger architecture while maintaining sufficient coherence. Other key questions involve more powerful quantum dynamics (e.g. "non-stoquastic" Hamiltonians) that cannot be efficiently classically simulated, advanced annealing protocols, tunneling, entanglement, error suppression, and others. The exploration of more-coherent quantum annealing reported here is one early step toward understanding the physics and the viability of quantum annealing as a computing resource.

ACKNOWLEDGMENT

We thank our Northrop Grumman colleagues for their many contributions; all collaborators on the QEO team; the team at MIT Lincoln Laboratory for very helpful discussions, collaboration, and support, including Jonilyn L. Yoder, Steve Weber, Jamie Kerman, David Kim, George Fitch, Bethany Niedzielski Huffman, Alexander Melville, Jovi Miloshi, Arjan Sevi, Danna Rosenberg, Gabriel Samach, Cyrus Hirjibehedin, Simon Gustavsson, and William D. Oliver.